\documentclass[12pt, prd, showpacs]{revtex4}
\usepackage{amssymb}
\usepackage{amsmath}

\begin{document}

\title{Near-horizon properties of trajectories with finite force relevant
for Ba\~{n}ados-Silk-West effect}
\author{H.V.Ovcharenko}
\affiliation{Department of Physics, V.N.Karazin Kharkov National University, 61022
Kharkov, Ukraine}
\affiliation{Institute of Theoretical Physics, Faculty of Mathematics and Physics,
Charles University, Prague, V Holesovickach 2, 180 00 Praha 8, Czech Republic}
\email{gregor_ovcharenko@outlook.com}
\author{O. B. Zaslavskii}
\affiliation{Department of Physics and Technology, Kharkov V.N. Karazin National
University, 4 Svoboda Square, Kharkov 61022, Ukraine}
\email{zaslav@ukr.net}

\begin{abstract}
According to the Banados-SIlk-West (BSW) effect, two particles moving
towards a black hole, can collide near the horizon with an unbounded energy
in the center of mass frame. This requires one of particles to have
fine-tuned parameters in such a way that the time component of generalized
momentum is zero $X=0$. Thus the existence of such trjectories is a
necessary condition for the BSW effect. However, it is insufficient since
the forward-in-time condition requires $X>0$ outside the horizon. We examine
this condition for different types of partricles and horizons and find
configurations for which the BSW effect is possible. In doing so, we take
into account a finite force of unspesified nature exerted on particles. It
includes relationships between numbers characterizing the rate with which
four-velocity, acceleration and metric functions change near the horizon.
For some aforementioned relations, parameters of a system control the sign
of $X$, in other cases they are required for $X$ to be real quantity. In the
simplest case of free particles the BSW effect for the Kerr or Kerr-Newman
black hole is impossible if a fine-tuned particle has a negative energy, so
in this sense combination of the Penrose process and the BSW effect is
forbidden.
\end{abstract}

\maketitle

\section{Introduction}

If two particles collide in the vicinity of a black hole, the energy $%
E_{c.m.}$ in the center of mass frame can grow unbounded, provided one of
particles has fine-tuned parameters \cite{ban}. This is the
Banados-Silk-West effect (BSW, after the names of its authors). Originally,
this was observed for free particles in the background of a rotating black
hole. However, counterpart of it exists also for static charged black holes 
\cite{jl} as well as for combination of electric charge and rotation \cite%
{bfh}. Instead of considering the electromagnetic force, one can scrutinize
the effect of a force as such, not specifying its nature. This helps to
understand the nature of the BSW effect, evaluate the role of gravitation
radiation in the context of the BSW effect \cite{insp} (at least
qualitatively, to the extent that it can be modeled by a force), take into
account the influence of surrounding medium, etc. For nonextremal and
extremal black holes, this was done in \cite{tz13}, \cite{tz14}. General
approach was developed in \cite{force23}.

The key ingredient of the BSW effect is the existence of a fine-tuned
trajectory for which the time component of generalized momentum $X=0$ on the
horizon. Meanwhile, there is also another condition that, to the best of our
knowledge, was not properly taken into account before. It consists in the
requirement $dX/dr>0$ in the immediate vicinity of the horizon to satisfy
the forward-in-time condition for a fine-tuned particle $X>0$ outside the
horizon. For a free moving particle, this is an obvious constraint on its
parameters but even in this case it leads to meaningful consequences
restricting possible scenarios of the BSW effect (see below). The situation
becomes nontrivial when a force acts on a particle since in some situations
the sign of $X$ outside the horizon cannot be chosen arbitrary and is
determined by dynamics, so one is led to the analysis of equations of
motion. But even if this sign can be imposed by hand, other constraints may
exist (for example, from the requirement that $X$ should be real quantity).
In other words, nontrivial interplay between kinematics and dynamics can
occur in the vicinity of the horizon as a factor selecting scenarios
appropriate for the BSW effect.

For example, for the Reissner-Nordsr\"{o}m metric $X=E-\frac{qQ}{r}$, where $%
E$ is the Killing energy, $q$ being particle charge, $Q$ a black hole mass.
If we choose parameters in such a way that $E=qQ/r_{h}$ on a horizon $%
r=r_{h} $, the fine-tuned (critical) trajectory does exist for $qQ>0$. If $%
qQ<0$, it still exists for $E<0$. However, this does not save the matter
since in an immediate vicinity of the horizon $X<0$ outside and this
destroys the BSW effect. In principle, similar restrictions should be valid
for neutral particles in the rotating black hole background that impose
constraint on the behavior of acceleration and other particle parameters on
the horizon.

The main goal of our work is to find conditions which have to be satisfied
by the external force to have $dX/dr>0$ in the vicinity of the horizon and
analyze for which types of particles this is in principle possible. To this
end, in Section \ref{sec_gen_set} we develop a general approach and for
particles moving in the equatorial plane we relate the rate of change $dX/dr$
with the external force. In Section \ref{sec_gen_case} we analyze all
possible distinct cases which give different expressions for $dX/dr$ and
summarize which conditions have to hold to make the desirable sign of $dX/dr$%
. In Section \ref{sec_dif_types} we apply the conditions developed in
Section \ref{sec_gen_case} to different types of particles and analyze for
which types this sign is controlled by an external force and how to keep it
positive by the action of an external force.

\section{General setup\label{sec_gen_set}}

We investigate the motion of particles in the background of a rotating black
hole which is described in the generalized Boyer-Lindquist coordinates $%
(t,r,\theta ,\varphi )$ by the metric:%
\begin{equation}
ds^{2}=-N^{2}dt^{2}+g_{\varphi \varphi }(dt-\omega d\varphi )^{2}+\frac{%
dr^{2}}{A}+g_{\theta \theta }d\theta ^{2},
\end{equation}%
where all metric coefficients do not depend on $t$ and $\varphi $. The
horizon is located at $r=r_{h}$ where $A(r_{h})=N(r_{h})=0$. Near the
horizon, we use a general expansion for the functions $N^{2}$, $A$ and $%
\omega $:%
\begin{equation}
N^{2}=\kappa _{p}v^{p}+o(v^{p}),\text{ \ \ }A=A_{q}v^{q}+o(v^{q}),
\label{an_exp}
\end{equation}%
\begin{equation}
\omega =\omega _{H}+\omega _{k}v^{k}+o(v^{k}),  \label{om_exp}
\end{equation}%
where $q,p$ and $k$ are numbers that characterize the rate of a change of
the metric functions near the horizon, and $v=r-r_{h}.$

If a particle is freely moving, the space-time symmetries with respect to $%
\partial _{t}$ and $\partial _{\varphi }$ impose conservation of the
corresponding components of the four-momentum: $mu_{t}=-E$, $mu_{\varphi }=L$%
. Here, $E\,\ $has the meaning of energy, $L$ being angular momentum, $m$
particle mass, $u^{\mu }$ the four-velocity. We assume the symmetry with
respect to the equatorial plane. In what follows, we restrict ourselves by
equatorial motion. Then, it follows from equations of motion that the
four-velocity of a free-falling particle can be written in the following
form:%
\begin{equation}
u^{\mu }=\left( \frac{\mathcal{X}}{N^{2}},\sigma \frac{\sqrt{A}}{N}P,0,\frac{%
\mathcal{L}}{g_{\varphi \varphi }}+\frac{\omega \mathcal{X}}{N^{2}}\right) ,
\label{4_vel}
\end{equation}%
where $\sigma =\pm 1$, $\mathcal{X}=\epsilon -\omega \mathcal{L}$, $\epsilon
=E/m$, $\mathcal{L}=L/m$ and $P$ is given by:%
\begin{equation}
P=\sqrt{\mathcal{X}^{2}-N^{2}\left( 1+\frac{\mathcal{L}^{2}}{g_{\varphi
\varphi }}\right) }.  \label{P}
\end{equation}

The forward-in-time condition states that $u^{t}>0$, whence%
\begin{equation}
\mathcal{X}>0  \label{ft}
\end{equation}%
outside the horizon. On the horizon itself $\mathcal{X}=0$ is admissible,
the corresponding particle (trajectory) is fine-tuned. If a force it does
not depend on $t$ and $\varphi $, the expression (\ref{4_vel}) retains its
form but with $E$ and $\mathcal{L}$ that can depend on $r$. In doing so, we
take advantages of some equations already derived in previous works,
especially in \cite{tz13}.

For description of motion near the horizon, one can introduce the tetrad
attached to an observer.\ To this end, it is convenient to use the so-called
zero angular momentum observer (ZAMO) \cite{72}. More precisely, we consider
such an observer for which $r=const$ and call it OZAMO ("orbital" ZAMO)
since both such an observer and fine-tuned particle do not cross the
horizon. As a result, components of acceleration in this frame remain
finite. Then, we rely on equation (114) from \cite{tz13}: 
\begin{equation}
\frac{d\mathcal{X}}{d\tau }=Na_{o}^{(t)}-\frac{d\omega }{d\tau }\mathcal{L}
\label{xt}
\end{equation}%
According to eq. (111) of \cite{tz13},%
\begin{equation}
a_{o}^{(t)}=\frac{N}{\mathcal{X}}\left[ \frac{\mathcal{L}}{\sqrt{g_{\phi }}}%
a_{o}^{(\varphi )}+\frac{u^{r}}{\sqrt{A}}a_{0}^{(r)}\right] \text{.}
\end{equation}%
Then, (\ref{xt}) is equivalent to%
\begin{equation}
\frac{d\mathcal{X}}{d\tau }=\frac{N^{2}}{\mathcal{X}}\left[ \frac{\mathcal{L}%
}{\sqrt{g_{\phi }}}a_{o}^{(\varphi )}+\frac{u^{r}}{\sqrt{A}}a_{0}^{(r)}%
\right] -\frac{d\omega }{d\tau }\mathcal{L}.
\end{equation}

Due to independence of $t$ and $\varphi $, we have%
\begin{equation}
\frac{d\mathcal{X}}{d\tau }=\frac{d\mathcal{X}}{dr}u^{r}\text{, }\frac{%
d\omega }{d\tau }=\frac{d\omega }{dr}u^{r},
\end{equation}%
whence%
\begin{equation}
\frac{d\mathcal{X}}{dr}=\frac{N^{2}}{\mathcal{X}}\left[ \frac{a_{0}^{(r)}}{%
\sqrt{A}}+\frac{\mathcal{L}}{\sqrt{g_{\phi }}u^{r}}a_{o}^{(\varphi )}\right]
-\mathcal{L}\omega ^{\prime }(r).
\end{equation}

Using (\ref{4_vel}), we can also rewrite it in the form%
\begin{equation}
\frac{d\mathcal{X}}{dr}=\frac{N^{2}}{\mathcal{X}\sqrt{A}}\left[
a_{0}^{(r)}+\sigma \frac{\mathcal{L}N}{\sqrt{g_{\phi }}P}a_{o}^{(\varphi )}%
\right] -\mathcal{L}\omega ^{\prime }(r)\text{.}  \label{x'}
\end{equation}

Now, we are interested in the question: when does the fine-tuned particle
become bad? By "bad" we imply that $\frac{d\mathcal{X}}{dr}(r_{h})<0$. Then,
it is impossible to have $\mathcal{X}(r_{h})=0$ since in a small vicinity of
the horizon in the outer region we would have $\mathcal{X}<0$ in
contradiction with the forward-in-time condition (\ref{ft})$.$

\subsection{Free particle: no Penrose effect for a fine-tuned one}

Let us consider a particular important case when a particle is free. We
assume $\omega ^{\prime }<0$ that is typical of asymptotically flat metrics
such as the Kerr and Kerr-Newman ones. We obtain that for a "bad" particle $%
\mathcal{L}<0$. In principle, a particle can have $E<0$ due to the
ergoregion and we can achieve $\mathcal{X}=0$ on the horizon due to $%
\mathcal{L}<0$. However, in a small vicinity of the horizon, (\ref{ft}) will
be violated. This means that near the horizon $\mathcal{L}>0$ for any
fine-tuned particle. In turn, this means that a fine-tuned particle must
have $E>0$.

In the Penrose process particle 0 decays to two fragments 1 and 2, where
particle 1 has a negative energy while particle 2 escapes to infinity thus
giving the energy gain \cite{pen}. It follows form what is said above that
for a fine-tuned particle the Penrose process is impossible.

\section{General case\label{sec_gen_case}}

Now, let us consider the most general case when some force acts on a
particle. Our aim is to find such conditions that near the horizon $\frac{d%
\mathcal{X}}{dr}>0$. To this end, we assume that near the horizon expansions
hold

\begin{eqnarray}
N^{2} &=&\kappa _{p}v^{p}+o(v^{p}),\text{ \ \ }A=A_{q}v^{q}+o(v^{q})\text{,}
\\
\omega &=&\omega _{H}+\omega _{k}v^{k}+o(v^{k}).
\end{eqnarray}

Additionally, we assume that near the horizon parameters of a fine-tuned
particle behave in such a way that

\begin{equation}
\mathcal{X}=X_{s}v^{s}+o(v^{s}),\text{ \ \ }\mathcal{L}%
=L_{H}+L_{b}v^{b}+o(v^{b})\text{,}
\end{equation}%
so $\mathcal{X}$ obeys the condition $\mathcal{X}(r_{h})=0$.

Acceleration near the horizon reads

\begin{eqnarray}
a_{o}^{(r)} &=&\left( a_{o}^{(r)}\right) _{n_{1}}v^{n_{1}}+o(v^{n_{1}}), \\
a_{o}^{(\varphi )} &=&\left( a_{o}^{(\varphi )}\right)
_{n_{2}}v^{n_{2}}+o(v^{n_{2}}).
\end{eqnarray}

For the four-velocity, using its normalization, one finds

\begin{equation}
u^{r}=-\frac{\sqrt{A}}{N}\sqrt{\mathcal{X}^{2}-N^{2}\left( 1+\frac{\mathcal{L%
}^{2}}{g_{\varphi }}\right) },
\end{equation}%
where minus sign is chosen because the particle is considered to be
infalling. Near-horizon behavior of this quantity was analyzed in \cite%
{force23}. Generally, this behavior may be described by the quantity $c$,
such that $u^{r}\approx \left( u^{r}\right) _{c}v^{c},$ where $c$ depends on
quantity $s.~$If $0\leq s\leq p/2,$ then $c=\frac{q-p}{2}+s.$ The case $%
s>p/2 $ is impossible, because the quantity $u^{r}$ would become complex.
However, if coefficients in expansion for $X$ are chosen in such a way that
they satisfy the condition

\begin{equation}
\mathcal{X}^{2}=N^{2}\left( 1+\frac{\mathcal{L}^{2}}{g_{\varphi }}\right) 
\text{ plus }v^{2c+p-q}\text{ terms,}
\end{equation}%
then $u^{r}\approx \left( u^{r}\right) _{c}v^{c}$ where $c$ may be any
value, higher then $q/2.$

Now let us analyze an expression (\ref{x'}) for $\frac{d\mathcal{X}}{dr}$

Substituting near-horizon expansions, one obtains

\begin{equation}
\frac{d\mathcal{X}}{dr}=sX_{s}v^{s-1}=\frac{\kappa _{p}}{X_{s}}\frac{\left(
a_{0}^{(r)}\right) _{n_{1}}}{\sqrt{A_{q}}}v^{n_{1}+p-\frac{q}{2}-s}+\frac{%
\kappa _{p}}{X_{s}}\frac{L_{H}\left( a_{o}^{(\phi )}\right) _{n_{2}}}{\sqrt{%
g_{\phi H}}\left( u^{r}\right) _{c}}v^{n_{2}+p-c-s}-L_{H}k\omega _{k}v^{k-1}.
\label{main_eq}
\end{equation}

\bigskip Thus we have 3 terms having the orders $n_{1}+p-\frac{q}{2}-s,$ $%
n_{2}+p-c-s$ and $k-1.$ We have to analyze all the cases when only one from
these terms is dominant, when 2 terms are of the same order and when all 3
terms are of the same order. To cover all these situations, let us analyze
them case by case.

\subsection{1-st term is dominant\label{3-1}}

For this to be true, one has to have%
\begin{eqnarray}
n_{2} &>&n_{1}+c-\frac{q}{2},  \label{n_2_rel_1st_term} \\
k &>&n_{1}+p-\frac{q}{2}+1-s  \label{k_rel_1st_term}
\end{eqnarray}

In this case, (\ref{main_eq}) becomes%
\begin{equation}
sX_{s}v^{s-1}=\frac{\kappa _{p}}{X_{s}}\frac{\left( a_{0}^{(r)}\right)
_{n_{1}}}{\sqrt{A_{q}}}v^{n_{1}+p-\frac{q}{2}-s}.
\end{equation}

From this equation it follows that $s$ has to satisfy the condition

\begin{equation}
n_{1}=2s+\frac{q}{2}-p-1
\end{equation}

Substituting this to (\ref{n_2_rel_1st_term}-\ref{k_rel_1st_term}), one
obtains that these conditions become%
\begin{eqnarray}
n_{2} &>&2s-p-1+c, \\
k &>&s
\end{eqnarray}

If it is so, then%
\begin{equation}
sX_{s}=\frac{\kappa _{p}}{X_{s}}\frac{\left( a_{0}^{(r)}\right) _{n_{1}}}{%
\sqrt{A_{q}}}\Rightarrow \left( X_{s}\right) ^{2}=\frac{\kappa _{p}}{s}\frac{%
\left( a_{0}^{(r)}\right) _{n_{1}}}{\sqrt{A_{q}}}.
\end{equation}

One sees that in this case we can define only $\left( X_{s}\right) ^{2},$
and thus a force does not control the sign of the $X_{s}.$ However, for $%
X_{s}$ to be real, one has to require $\left( a_{0}^{(r)}\right) _{n_{1}}>0.$

\subsection{2-nd term is dominant\label{3-2}}

For this to be true, one has to have%
\begin{eqnarray}
n_{1} &>&n_{2}+\frac{q}{2}-c,  \label{n_1_rel_2nd_term} \\
k &>&n_{2}+p-c+1-s.  \label{k_rel_2nd_term}
\end{eqnarray}

In this case, (\ref{main_eq}) becomes%
\begin{equation}
sX_{s}v^{s-1}=\frac{\kappa _{p}}{X_{s}}\frac{L_{H}\left( a_{o}^{(\phi
)}\right) _{n_{2}}}{\sqrt{g_{\phi H}}\left( u^{r}\right) _{c}}%
v^{n_{2}+p-c-s}.
\end{equation}

From this equation it follows that $s$ has to satisfy the condition 
\begin{equation}
n_{2}=2s+c-p-1.
\end{equation}

Substituting this to (\ref{n_1_rel_2nd_term}-\ref{k_rel_2nd_term}), one gets%
\begin{eqnarray}
n_{1} &>&2s+\frac{q}{2}-p-1, \\
k &>&s.
\end{eqnarray}

If it is so, then%
\begin{equation}
sX_{s}=\frac{\kappa _{p}}{X_{s}}\frac{L_{H}\left( a_{o}^{(\phi )}\right)
_{n_{2}}}{\sqrt{g_{\phi H}}\left( u^{r}\right) _{c}}\Rightarrow \left(
X_{s}\right) ^{2}=\frac{\kappa _{p}}{s}\frac{L_{H}\left( a_{o}^{(\phi
)}\right) _{n_{2}}}{\sqrt{g_{\phi H}}\left( u^{r}\right) _{c}}.
\end{equation}

One sees that in this case we can define only $\left( X_{s}\right) ^{2},$
and thus a force does not control the sign of the $X_{s}.$ However, for $%
X_{s}$ to be real, one has to require $\frac{L_{H}\left( a_{o}^{(\phi
)}\right) _{n_{2}}}{\left( u^{r}\right) _{c}}>0.$ As for infalling particles 
$\left( u^{r}\right) _{c}<0$, signs of angular momentum and angular
components of acceleration have to be different.

\subsection{3-rd term is dominant\label{3-3}}

For this to be true, one has to have%
\begin{eqnarray}
n_{1} &>&k-1+\frac{q}{2}-p+s,  \label{n_1_rel_3rd_term} \\
n_{2} &>&k-1+c-p+s.  \label{n_2_rel_3rd_term}
\end{eqnarray}

In this case, (\ref{main_eq}) becomes%
\begin{equation}
sX_{s}v^{s-1}=-L_{H}k\omega _{k}v^{k-1}.
\end{equation}

From this equation it follows that $s$ has to satisfy the condition$.$%
\begin{equation}
s=k.
\end{equation}

Substituting this to (\ref{n_1_rel_3rd_term}-\ref{n_2_rel_3rd_term}), one
gets%
\begin{eqnarray}
n_{1} &>&2s+\frac{q}{2}-p-1, \\
n_{2} &>&2s+c-p-1.
\end{eqnarray}

If it is so, then%
\begin{equation}
X_{s}=-L_{H}\omega _{k}.
\end{equation}

$X_{s}$ will be positive if $L_{H}\omega _{k}<0$ only$.$ This condition is
the same as for the case of freely-moving particles.

\subsection{1-st and 2-nd terms are dominant\label{3-4}}

For this to be true, one has to have%
\begin{eqnarray}
n_{2} &=&n_{1}+c-\frac{q}{2},  \label{n_2_rel_12_term} \\
k &>&n_{1}+p-\frac{q}{2}+1-s.  \label{k_rel_12_term}
\end{eqnarray}

In this case, (\ref{main_eq}) becomes%
\begin{equation}
sX_{s}v^{s-1}=\frac{\kappa _{p}}{X_{s}}\left[ \frac{\left(
a_{0}^{(r)}\right) _{n_{1}}}{\sqrt{A_{q}}}+\frac{L_{H}\left( a_{o}^{(\phi
)}\right) _{n_{2}}}{\sqrt{g_{\phi H}}\left( u^{r}\right) _{c}}\right]
v^{n_{1}+p-\frac{q}{2}-s}.
\end{equation}

From this equation it follows that $s$ has to satisfy the condition%
\begin{equation}
n_{1}=2s+\frac{q}{2}-p-1.
\end{equation}

Substituting this to (\ref{n_2_rel_12_term}-\ref{k_rel_12_term}), one gets%
\begin{eqnarray}
n_{2} &=&2s+c-p-1, \\
k &>&s.
\end{eqnarray}

If it is so, then%
\begin{equation}
sX_{s}=\frac{\kappa _{p}}{X_{s}}\left[ \frac{\left( a_{0}^{(r)}\right)
_{n_{1}}}{\sqrt{A_{q}}}+\frac{L_{H}\left( a_{o}^{(\phi )}\right) _{n_{2}}}{%
\sqrt{g_{\phi H}}\left( u^{r}\right) _{c}}\right] \Rightarrow \left(
X_{s}\right) ^{2}=\frac{\kappa _{p}}{s}\left[ \frac{\left(
a_{0}^{(r)}\right) _{n_{1}}}{\sqrt{A_{q}}}+\frac{L_{H}\left( a_{o}^{(\phi
)}\right) _{n_{2}}}{\sqrt{g_{\phi H}}\left( u^{r}\right) _{c}}\right] .
\end{equation}

One sees that in this case we can define only $\left( X_{s}\right) ^{2},$
and thus we cannot control the sign of the $X_{s}.$ However, for $X_{s}$ to
be real, one has to require $\frac{\left( a_{0}^{(r)}\right) _{n_{1}}}{\sqrt{%
A_{q}}}+\frac{L_{H}\left( a_{o}^{(\phi )}\right) _{n_{2}}}{\sqrt{g_{\phi H}}%
\left( u^{r}\right) _{c}}>0.$

\subsection{All three terms are of the same order\label{3-5}}

Now let us analyze this case. We postpone analysis of 2 additional cases
when 1-st and 3-rd terms and 2-nd and 3-rd terms are dominant. Motivation
for this is that these cases could be obtained from the case when all 3
terms are of the same order after taking several parameters to zero. For
this to be true, one has to have%
\begin{eqnarray}
n_{2} &=&n_{1}+c-\frac{q}{2},  \label{n_2_rel_123_term} \\
n_{1} &=&k+\frac{q}{2}-p+s-1.  \label{n_1_rel_123_term}
\end{eqnarray}

In this case, (\ref{main_eq}) becomes%
\begin{equation}
sX_{s}v^{s-1}=\left( \frac{\kappa _{p}}{X_{s}}\left[ \frac{\left(
a_{0}^{(r)}\right) _{n_{1}}}{\sqrt{A_{q}}}+\frac{L_{H}\left( a_{o}^{(\phi
)}\right) _{n_{2}}}{\sqrt{g_{\phi H}}\left( u^{r}\right) _{c}}\right]
-L_{H}k\omega _{k}\right) v^{n_{1}+p-\frac{q}{2}-s}.
\end{equation}

From this equation it follows that $s$ has to satisfy the condition%
\begin{equation}
n_{1}=2s+\frac{q}{2}-p-1.
\end{equation}

Substituting this to (\ref{n_2_rel_123_term}-\ref{n_1_rel_123_term}), one
gets%
\begin{eqnarray}
n_{2} &=&2s+c-p-1, \\
k &=&s.
\end{eqnarray}

If it is so, then%
\begin{equation}
sX_{s}=\frac{\kappa _{p}}{X_{s}}\left[ \frac{\left( a_{0}^{(r)}\right)
_{n_{1}}}{\sqrt{A_{q}}}+\frac{L_{H}\left( a_{o}^{(\phi )}\right) _{n_{2}}}{%
\sqrt{g_{\phi H}}\left( u^{r}\right) _{c}}\right] -L_{H}k\omega _{k}.
\end{equation}

Multiplying by $X_{s},$ one obtains%
\begin{equation}
s\left( X_{s}\right) ^{2}+kL_{H}\omega _{k}X_{s}-\kappa _{p}\left[ \frac{%
\left( a_{0}^{(r)}\right) _{n_{1}}}{\sqrt{A_{q}}}+\frac{L_{H}\left(
a_{o}^{(\phi )}\right) _{n_{2}}}{\sqrt{g_{\phi H}}\left( u^{r}\right) _{c}}%
\right] =0.
\end{equation}

Solving this equation, we find%
\begin{equation}
X_{s}=-\frac{kL_{H}\omega _{k}}{2s}\pm \frac{\sqrt{D}}{2s}\text{,}
\end{equation}

where 
\begin{equation}
D=(kL_{H}\omega _{k})^{2}+4s\kappa _{p}\left[ \frac{\left(
a_{0}^{(r)}\right) _{n_{1}}}{\sqrt{A_{q}}}+\frac{L_{H}\left( a_{o}^{(\phi
)}\right) _{n_{2}}}{\sqrt{g_{\phi H}}\left( u^{r}\right) _{c}}\right] .
\end{equation}

To analyze what signs these roots can get, let us introduce%
\begin{equation}
a=\frac{kL_{H}\omega _{k}}{2s},\text{ \ \ }b=\frac{\kappa _{p}}{s}\left[ 
\frac{\left( a_{0}^{(r)}\right) _{n_{1}}}{\sqrt{A_{q}}}+\frac{L_{H}\left(
a_{o}^{(\phi )}\right) _{n_{2}}}{\sqrt{g_{\phi H}}\left( u^{r}\right) _{c}}%
\right] .
\end{equation}

Thus the solution under discussion becomes%
\begin{equation}
X_{s}=-a\pm \sqrt{a^{2}+b}.
\end{equation}

If $b>0,$ then, independently on the sign of $a$, one of the roots is
positive. But if $b<0,$ then one of the roots is positive only if $a<0.$
These conditions mean that if%
\begin{equation}
\frac{\left( a_{0}^{(r)}\right) _{n_{1}}}{\sqrt{A_{q}}}+\frac{L_{H}\left(
a_{o}^{(\phi )}\right) _{n_{2}}}{\sqrt{g_{\phi H}}\left( u^{r}\right) _{c}}>0
\end{equation}

then positive $X_{s}$ exists independently on $L_{H}\omega _{k}.$ However,
if 
\begin{equation}
\frac{\left( a_{0}^{(r)}\right) _{n_{1}}}{\sqrt{A_{q}}}+\frac{L_{H}\left(
a_{o}^{(\phi )}\right) _{n_{2}}}{\sqrt{g_{\phi H}}\left( u^{r}\right) _{c}}<0
\end{equation}

then positive $X_{s}$ exists only if $L_{H}\omega _{k}<0.$ Additionally, we
have to mention that if 
\begin{equation}
\frac{\left( a_{0}^{(r)}\right) _{n_{1}}}{\sqrt{A_{q}}}+\frac{L_{H}\left(
a_{o}^{(\phi )}\right) _{n_{2}}}{\sqrt{g_{\phi H}}\left( u^{r}\right) _{c}}=0
\end{equation}

then there is only one root%
\begin{equation}
X_{s}=-2a
\end{equation}

This root is positive if $L_{H}\omega _{k}<0$ only$.$

\subsection{1-st and 3-rd terms are dominant\label{3-6}}

For this to be true, one has to have%
\begin{eqnarray}
n_{1} &=&k+\frac{q}{2}-p+s-1,  \label{n_1_rel_13_term} \\
n_{2} &>&k+c-1-p+s  \label{n_2_rel_13_term}
\end{eqnarray}

In this case, (\ref{main_eq}) becomes%
\begin{equation}
sX_{s}v^{s-1}=\left( \frac{\kappa _{p}}{X_{s}}\frac{\left(
a_{0}^{(r)}\right) _{n_{1}}}{\sqrt{A_{q}}}-L_{H}k\omega _{k}\right)
v^{n_{1}+p-\frac{q}{2}-s}.
\end{equation}

From this equation it follows that $s$ has to satisfy condition 
\begin{equation}
n_{1}=2s+\frac{q}{2}-p-1.
\end{equation}%
Substituting this to (\ref{n_1_rel_13_term}-\ref{n_2_rel_13_term}), one gets%
\begin{eqnarray}
n_{2} &>&2s+c-p-1, \\
k &=&s.
\end{eqnarray}%
If it is so, then%
\begin{equation}
sX_{s}=\frac{\kappa _{p}}{X_{s}}\frac{\left( a_{0}^{(r)}\right) _{n_{1}}}{%
\sqrt{A_{q}}}-L_{H}k\omega _{k}
\end{equation}

Multiplying by $X_{s}$ (only if $X_{s}\neq 0$)$,$ one obtains%
\begin{equation}
s\left( X_{s}\right) ^{2}+kL_{H}\omega _{k}X_{s}-\kappa _{p}\frac{\left(
a_{0}^{(r)}\right) _{n_{1}}}{\sqrt{A_{q}}}=0
\end{equation}

Solution of this equation is%
\begin{equation*}
X_{s}=-a\pm \sqrt{a^{2}+b}
\end{equation*}

where%
\begin{equation}
a=\frac{kL_{H}\omega _{k}}{2s},\text{ \ \ }b=\frac{\kappa _{p}}{s}\frac{%
\left( a_{0}^{(r)}\right) _{n_{1}}}{\sqrt{A_{q}}}
\end{equation}

One sees that this solution is similar to the case when all 3 roots are of
the same order, but with a slight change of coefficient $b.$ Following the
lines, we see, that if 
\begin{equation}
\frac{\left( a_{0}^{(r)}\right) _{n_{1}}}{\sqrt{A_{q}}}>0\text{,}
\end{equation}

then positive $X_{s}$ exists independently on $L_{H}\omega _{k}.$ However,
if 
\begin{equation}
\frac{\left( a_{0}^{(r)}\right) _{n_{1}}}{\sqrt{A_{q}}}<0\text{,}
\end{equation}

then positive $X_{s}$ exists only if $L_{H}\omega _{k}<0.$ Additionally, we
have to mention that if 
\begin{equation}
\frac{\left( a_{0}^{(r)}\right) _{n_{1}}}{\sqrt{A_{q}}}=0,
\end{equation}

then there is only one root%
\begin{equation}
X_{s}=-2a
\end{equation}

This root is positive only if $L_{H}\omega _{k}<0.$

\subsection{2-nd and 3-rd terms are dominant\label{3-7}}

For this to be true, one has to have%
\begin{eqnarray}
n_{2} &=&k+c-1-p+s,  \label{n_2_rel_23_term} \\
n_{1} &>&k+\frac{q}{2}-p+s-1.  \label{k_rel_23_term}
\end{eqnarray}

In this case, (\ref{main_eq}) becomes%
\begin{equation}
sX_{s}v^{s-1}=\left( \frac{\kappa _{p}}{X_{s}}\frac{L_{H}\left( a_{o}^{(\phi
)}\right) _{n_{2}}}{\sqrt{g_{\phi H}}\left( u^{r}\right) _{c}}-L_{H}k\omega
_{k}\right) v^{n_{2}+p-c-s}.
\end{equation}

From this equation it follows that $s$ has to satisfy the condition 
\begin{equation}
n_{2}=2s+c-p-1.
\end{equation}

Substituting this to (\ref{n_2_rel_23_term}-\ref{k_rel_23_term}), one gets%
\begin{eqnarray}
n_{1} &>&2s+\frac{q}{2}-p-1, \\
k &=&s.
\end{eqnarray}

If it is so, then%
\begin{equation}
sX_{s}=\frac{\kappa _{p}}{X_{s}}\frac{L_{H}\left( a_{o}^{(\phi )}\right)
_{n_{2}}}{\sqrt{g_{\phi H}}\left( u^{r}\right) _{c}}-L_{H}k\omega _{k}.
\end{equation}

Multiplying by $X_{s}$ (only if $X_{s}\neq 0$)$,$ one obtains%
\begin{equation}
s\left( X_{s}\right) ^{2}+kL_{H}\omega _{k}X_{s}-\kappa _{p}\frac{%
L_{H}\left( a_{o}^{(\phi )}\right) _{n_{2}}}{\sqrt{g_{\phi H}}\left(
u^{r}\right) _{c}}=0
\end{equation}

Solution of this equation is%
\begin{equation*}
X_{s}=-a\pm \sqrt{a^{2}+b},
\end{equation*}

where%
\begin{equation}
a=\frac{kL_{H}\omega _{k}}{2s},\text{ \ \ }b=\frac{\kappa _{p}}{s}\frac{%
L_{H}\left( a_{o}^{(\phi )}\right) _{n_{2}}}{\sqrt{g_{\phi H}}\left(
u^{r}\right) _{c}}.
\end{equation}

One sees that this solution is similar to the case when all 3 roots are of
the same order, but with a slight change of coefficient $b.$ Following the
same analysis as in the previous case, we see that if 
\begin{equation}
\frac{L_{H}\left( a_{o}^{(\phi )}\right) _{n_{2}}}{\sqrt{g_{\phi H}}\left(
u^{r}\right) _{c}}>0,
\end{equation}

then positive $X_{s}$ exists independently on $L_{H}\omega _{k}.$ However,
if 
\begin{equation}
\frac{L_{H}\left( a_{o}^{(\phi )}\right) _{n_{2}}}{\sqrt{g_{\phi H}}\left(
u^{r}\right) _{c}}<0
\end{equation}

then positive $X_{s}$ exists only if $L_{H}\omega _{k}<0.$ Additionally, we
have to mention that if 
\begin{equation}
\frac{L_{H}\left( a_{o}^{(\phi )}\right) _{n_{2}}}{\sqrt{g_{\phi H}}\left(
u^{r}\right) _{c}}=0
\end{equation}

then there is only one root%
\begin{equation}
X_{s}=-2a
\end{equation}

This root is positive only if $L_{H}\omega _{k}<0.$

\begin{table}[tbp]
\begin{tabular}{|c|c|c|}
\hline
Dominant terms & Restrictions on $n_{1},$ $n_{2},$ $k$ & Conditions which
have to hold for $X_{s}>0$ \\ \hline
1-st & 
\begin{tabular}{l}
$n_{1}=2s+\frac{q}{2}-p-1$ \\ 
$n_{2}>2s+c-p-1$ \\ 
$k>s$%
\end{tabular}
& absent, but $\left( a_{0}^{(r)}\right) _{n_{1}}>0$ for $X_{s}$ to be real
\\ \hline
2-nd & 
\begin{tabular}{l}
$n_{1}>2s+\frac{q}{2}-p-1$ \\ 
$n_{2}=2s+c-p-1$ \\ 
$k>s$%
\end{tabular}
& absent, but $\frac{L_{H}\left( a_{o}^{(\phi )}\right) _{n_{2}}}{\left(
u^{r}\right) _{c}}>0$ for $X_{s}$ to be real \\ \hline
3-rd & 
\begin{tabular}{l}
$n_{1}>2s+\frac{q}{2}-p-1$ \\ 
$n_{2}>2s+c-p-1$ \\ 
$k=s$%
\end{tabular}
& $L_{H}\omega _{k}<0$ \\ \hline
1-st and 2-nd & 
\begin{tabular}{l}
$n_{1}=2s+\frac{q}{2}-p-1$ \\ 
$n_{2}=2s+c-p-1$ \\ 
$k>s$%
\end{tabular}
& absent, but $\ \frac{\left( a_{0}^{(r)}\right) _{n_{1}}}{\sqrt{A_{q}}}+%
\frac{L_{H}\left( a_{o}^{(\phi )}\right) _{n_{2}}}{\sqrt{g_{\phi H}}\left(
u^{r}\right) _{c}}>0$ for $X_{s}$ to be real \\ \hline
1-st and 3-rd & 
\begin{tabular}{l}
$n_{1}=2s+\frac{q}{2}-p-1$ \\ 
$n_{2}>2s+c-p-1$ \\ 
$k=s$%
\end{tabular}
& $\frac{\left( a_{0}^{(r)}\right) _{n_{1}}}{\sqrt{A_{q}}}>0$ or if $\frac{%
\left( a_{0}^{(r)}\right) _{n_{1}}}{\sqrt{A_{q}}}\leq 0$ and $L_{H}\omega
_{k}<0$ \\ \hline
2-nd and 3-rd & 
\begin{tabular}{l}
$n_{1}>2s+\frac{q}{2}-p-1$ \\ 
$n_{2}=2s+c-p-1$ \\ 
$k=s$%
\end{tabular}
& $\frac{L_{H}\left( a_{o}^{(\phi )}\right) _{n_{2}}}{\sqrt{g_{\phi H}}%
\left( u^{r}\right) _{c}}>0$ or $\frac{L_{H}\left( a_{o}^{(\phi )}\right)
_{n_{2}}}{\sqrt{g_{\phi H}}\left( u^{r}\right) _{c}}\leq 0$ and $L_{H}\omega
_{k}<0$ \\ \hline
1-st, 2-nd and 3-rd & 
\begin{tabular}{l}
$n_{1}=2s+\frac{q}{2}-p-1$ \\ 
$n_{2}=2s+c-p-1$ \\ 
$k=s$%
\end{tabular}
& $b>0$ or $b\leq 0$ and $L_{H}\omega _{k}<0$ \\ \hline
\end{tabular}%
\caption{ Table showing which conditions have to hold for different terms to
be dominant in the expression for $\frac{dX}{dr}$ and under which conditions
this quantity is positive. Here, the 1st term means the term with the radial
acceleration in eq. (12), the 2nd one corresponds to coupling between the
angular acceleration and angular momentum there, the 3d term does not
contain acceleration and arises entirely due to the angular momentum. We
denoted $b=\frac{\protect\kappa _{p}}{s}\left[ \frac{\left(
a_{0}^{(r)}\right) _{n_{1}}}{\protect\sqrt{A_{q}}}+\frac{L_{H}\left( a_{o}^{(%
\protect\phi )}\right) _{n_{2}}}{\protect\sqrt{g_{\protect\phi H}}\left(
u^{r}\right) _{c}}\right] .$}
\label{tab_1}
\end{table}

We summarize all the possible cases in Table \ref{tab_1}.

\section{Particular cases}

\subsection{Radial acceleration}

Let us also derive general relations in the case when acceleration has only
the radial component $a_{o}^{(r)}\neq 0$, whereas $a_{o}^{(\phi )}=0.$
Behavior of $\frac{d\mathcal{X}}{dr}$ in this case can be found using the
results we previously obtained. Indeed, if $a_{o}^{(\phi )}=0$ then (\ref%
{main_eq}) becomes

\begin{equation}
sX_{s}v^{s-1}=\frac{\kappa _{p}}{X_{s}}\frac{\left( a_{0}^{(r)}\right)
_{n_{1}}}{\sqrt{A_{q}}}v^{n_{1}+p-\frac{q}{2}-s}-L_{H}k\omega _{k}v^{k-1}.
\end{equation}

Thus, we have only 2 terms, which we have to compare. If $n_{1}<k+\frac{q}{2}%
+s-p-1,$ then the first term is dominant. This case was already analyzed and
is presented by the first line in Table \ref{tab_1}.

If $n_{1}=k+\frac{q}{2}+s-p-1,$ then both terms are of the same order and
this case corresponds to the 5-th line in Table \ref{tab_1}.

If $n_{1}>k+\frac{q}{2}+s-p-1,$ then the second term is dominant and this
case corresponds to the 3-rd line in Table \ref{tab_1}.

\subsection{Angular acceleration}

In the opposite case $a_{o}^{(r)}=0$, $a_{o}^{(\phi )}\neq 0.$ Then, (\ref%
{main_eq}) becomes%
\begin{equation*}
sX_{s}v^{s-1}=\frac{\kappa _{p}}{X_{s}}\frac{L_{H}\left( a_{o}^{(\phi
)}\right) _{n_{2}}}{\sqrt{g_{\phi H}}\left( u^{r}\right) _{c}}%
v^{n_{2}+p-c-s}-L_{H}k\omega _{k}v^{k-1}.
\end{equation*}

If $n_{2}=n_{2}^{\ast }\equiv k+\frac{q}{2}-1-p+s,$ the 1-st terms is
dominant and this corresponds to the second line in Table \ref{tab_1}. If $%
n_{2}=n_{2}^{\ast }$, both terms are of the same order and this corresponds
to the 6-th line in Table \ref{tab_1}. If $n_{2}>n_{2}^{\ast }$, the second
term is dominant and this corresponds to the 3-rd line in Table \ref{tab_1}.

The case of pure angular accelertaion was already considered in Sec. V C of 
\cite{tz13} for a particular case of extremal black holes. It corresponds to 
$p=q=2$, $b=1=k=n_{2}$, so $n_{2}^{\ast }=0$ and $n_{2}>n_{2}^{\ast }$
(where $n_{2}^{\ast }$ is the value of the degree $n_{2}$ from the article 
\cite{tz13}).

\subsection{Static space-time}

For static spacetimes there is no $\omega $ and (\ref{main_eq}) becomes

\begin{equation}
sX_{s}v^{s-1}=\frac{\kappa _{p}}{X_{s}}\frac{\left( a_{0}^{(r)}\right)
_{n_{1}}}{\sqrt{A_{q}}}v^{n_{1}+p-\frac{q}{2}-s}+\frac{\kappa _{p}}{X_{s}}%
\frac{L_{H}\left( a_{o}^{(\phi )}\right) _{n_{2}}}{\sqrt{g_{\phi H}}\left(
u^{r}\right) _{c}}v^{n_{2}+p-c-s}.
\end{equation}

As one can see, that terms at RHS are the same as the first and second terms
in (\ref{main_eq}). Then, the results of subsection \ref{3-4} apply.
According to Table \ref{tab_1}, for all these cases a force does not control
the sign of $X_{s}.$ In \cite{zas-jetp} the particular case of the
Schwarzschild black hole with a radial force acting on a particle was
considered. There, the sign of acceleration was chosen by hand to ensure the
existence of critical trajectories but the question about the sign of $dX/dr$
was not posed. Now, we are making the second step desribing this issue.

\section{Different types of particles\label{sec_dif_types}}

Now let us consider different types of particles and how the possibility of
having positive $dX/dr$ is related to the type of a particle. For this we
have to review main properties of different types of particles. As was said
in Section \ref{sec_gen_case}, they differ by the relation between numbers $%
s $ and $c.$ Namely, for subcritical particles $0<s<\frac{p}{2},$ $c=s+\frac{%
q-p}{2},$ for critical $s=\frac{p}{2},$ $c=\frac{q}{2},$ for ultracritical $%
s=\frac{p}{2},$ $c>\frac{q}{2}$ (see Table 1 in \cite{force23}). Now let us
apply this for to the analysis of different cases to elucidate when a force
controls the sign of $dX/dr$. As follows from the Table \ref{tab_1}, this is
possible only in cases when either only 3-rd term is dominant, or if 1-st
and 2-nd, 1-st and 3-rd and 2-nd and 3-rd terms are dominant, or if all 3
terms are of the same order. Thus we have to consider which conditions we
obtain for each of these cases for different types of particles. Also note
that one can additionally add requirement of finiteness of external forces
(namely, requiring $n_{1,2}\geq 0$), but extensive analysis of conditions
that have to meet to make force finite was already done in \cite{force23}
(see Tables IV, V and VI in there), so we will not repeat them here. Only in
next Section we will provide this analysis for different types of horizons
and show that it correlates with the results in \cite{force23}.

\subsection{3-rd term is dominant}

3-rd term is dominant if $n_{1}>2s+\frac{q}{2}-p-1,$ $n_{1}>2s+c-p-1$ and $%
k=s.$

For subcritical particles $c=s+\frac{q-p}{2},$ thus conditions for $n_{1,2}$
become:%
\begin{eqnarray}
n_{1} &>&2s+\frac{q}{2}-p-1 \\
n_{2} &>&2s+\frac{q}{2}-p-1+\left( s-\frac{p}{2}\right)
\end{eqnarray}

As for subcritical particles $s<\frac{p}{2},$ the lower bound for $n_{2}$
has to be smaller then the lower bound for $n_{1}.$ However, the conditions
themselves do not tell which number is greater, so there may exist any
relation between them. However, it is obvious that, as for subcritical
particles $0<s,$ independently on the exact value of $s$ holds $n_{1}>\frac{q%
}{2}-p-1,$ $n_{2}>\frac{q}{2}-\frac{3p}{2}-1.$ Condition $k=s$ gives us for
subcritical particles $0<k=s<\frac{p}{2}.$ Thus, summarizing

\textit{For subcritical particles relation between }$n_{1}$\textit{\ and }$%
n_{2}$\textit{\ is arbitrary, but }$n_{1}>\frac{q}{2}-p-1,n_{2}>\frac{q}{2}-%
\frac{3p}{2}-1.$\textit{\ Additionally }$0<k=s<\frac{p}{2}$

For critical particles $s=\frac{p}{2}$ and $c=\frac{q}{2}.$ Substituting
this to conditions on $n_{1,2},$ one gets%
\begin{eqnarray}
n_{1} &>&\frac{q}{2}-1 \\
n_{2} &>&\frac{q}{2}-1
\end{eqnarray}

One sees, that the lower bounds for both $n_{1,2}$ are the same, but
conditions do not restrict which number has to be greater. Additionally $k=s=%
\frac{p}{2}.$ Summarizing

\textit{For critical particles relation between }$n_{1}$\textit{\ and }$%
n_{2} $\textit{\ is arbitrary, but }$n_{1,2}>\frac{q}{2}-1.$\textit{\
Additionally }$k=\frac{p}{2}.$

For ultracritical particles $s=\frac{p}{2},$ $c>\frac{q}{2}.$ Conditions for 
$n_{1,2}$ give

\begin{eqnarray}
n_{1} &>&\frac{q}{2}-1 \\
n_{2} &>&c-1
\end{eqnarray}

In this case, lower bound for $n_{1}$ is smaller, but the relation between
these numbers is not restricted. Additionally, $k=s=\frac{p}{2}.$

\textit{For ultracritical particles relation between }$n_{1}$\textit{\ and }$%
n_{2}$\textit{\ is arbitrary, but }$n_{1}>\frac{q}{2}-1,$ $n_{2}>c-1.$%
\textit{\ Additionally, }$k=\frac{p}{2}.$

\subsection{1-st and 3-rd terms are dominant}

1-st and 2-nd terms are dominant if $n_{1}=2s+\frac{q}{2}-p-1,$ $%
n_{2}>2s+c-p-1,$ $k=s.$

For subcritical particles $c=s+\frac{q-p}{2},$ thus conditions for $n_{1,2}$
become:

\begin{eqnarray}
n_{1} &=&2s+\frac{q}{2}-p-1 \\
n_{2} &>&2s+\frac{q}{2}-p-1+\left( s-\frac{p}{2}\right)
\end{eqnarray}

From these conditions we see that $n_{2}>n_{1}+\left( s-\frac{p}{2}\right) .$
As $s<\frac{p}{2},$ we see that from this condition does not follow what is
greater: $n_{1}$ or $n_{2},$ but for them has to follow $n_{1}>\frac{q}{2}%
-p-1,$ $n_{2}>\frac{q}{2}-\frac{3p}{2}-1.$ Additionally $0<k=s<\frac{p}{2}$.
Thus summarizing

\textit{For subcritical particles relation between }$n_{1}$\textit{\ and }$%
n_{2}$\textit{\ is arbitrary, but }$n_{1}>\frac{q}{2}-p-1,$ $n_{2}>\frac{q}{2%
}-\frac{3p}{2}-1.$\textit{\ Additionally }$0<k=s<\frac{p}{2}.$

For critical particles $s=\frac{p}{2}$ and $c=\frac{q}{2}$ and conditions
for $n_{1,2}$ become%
\begin{eqnarray}
n_{1} &=&\frac{q}{2}-1 \\
n_{2} &>&\frac{q}{2}-1
\end{eqnarray}

Thus we see, that for them holds $n_{2}>n_{1}.$ Additionally, $k=\frac{p}{2}%
. $ Thus summarizing we have

\textit{For critical particles }$n_{2}>n_{1}=\frac{q}{2}-1.$\textit{\
Additionally, }$k=p/2.$

For ultracritical $s=\frac{p}{2},$ $c>\frac{q}{2}.$ Thus conditions for $%
n_{1,2}$ becomes

\begin{eqnarray}
n_{1} &=&\frac{q}{2}-1 \\
n_{2} &>&c-1
\end{eqnarray}

As $c>\frac{q}{2},$ we see that $n_{2}>n_{1}.$ Additionally $k=\frac{p}{2}.$
Thus summarizing we have

\textit{For ultracritical particles }$n_{2}>n_{1}=\frac{q}{2}-1.$\textit{\
Additionally, }$k=p/2.$

\subsection{2-nd and 3-rd terms are dominant}

2-nd and 3-rd terms are dominant if $n_{1}>2s+\frac{q}{2}-p-1,$ $%
n_{2}=2s+c-p-1,$ $k=s.$

For subcritical particles the conditions for $n_{1,2}$ become

\begin{eqnarray}
n_{1} &>&2s+\frac{q}{2}-p-1, \\
n_{2} &=&2s+\frac{q}{2}-p-1+\left( s-\frac{p}{2}\right) .
\end{eqnarray}

Combining these conditions we have $n_{1}>n_{2}+\left( \frac{p}{2}-s\right)
. $ As $s<\frac{p}{2}$ we see that $n_{1}>n_{2}.$ Additionally, $0<k=s<\frac{%
p}{2}.$ Thus summarizing, we have

\textit{for subcritical particles }$n_{1}>n_{2}$\textit{\ and }$n_{1}>\frac{q%
}{2}-p-1,$ $n_{2}>\frac{q}{2}-\frac{3p}{2}-1.$\textit{\ Additionally, }$%
0<k=s<\frac{p}{2}.$

For critical particles conditions for $n_{1,2}$ become%
\begin{eqnarray}
n_{1} &>&\frac{q}{2}-1. \\
n_{2} &=&\frac{q}{2}-1.
\end{eqnarray}

One can see that $n_{1}>n_{2}.$ Additionally, $k=\frac{p}{2}.$ Thus
summarizing,

\textit{for critical particles }$n_{1}>n_{2}$\textit{\ and }$n_{1}>\frac{q}{2%
}-1,$ $n_{2}=\frac{q}{2}-1.$\textit{\ Additionally, }$k=s=\frac{p}{2}.$

For ultracritical particles $s=\frac{p}{2}$ and $c>\frac{q}{2}$, so the
conditions for $n_{1,2}$ become

\begin{eqnarray}
n_{1} &>&\frac{q}{2}-1, \\
n_{2} &=&c-1.
\end{eqnarray}

As $c>\frac{q}{2},$ $n_{2}>\frac{q}{2}-1$ and we see that the relation
between $n_{1}$ and $n_{2}$ can be arbitrary. Additionally, $k=s=\frac{p}{2}%
. $ Thus summarizing,

\textit{for ultracritical particles relation between }$n_{1}$\textit{\ and }$%
n_{2}$\textit{\ is arbitrary, but }$n_{1,2}>\frac{q}{2}-1.$\textit{\
Additionally }$k=\frac{p}{2}.$

\subsection{All 3 terms are of the same order}

All 3 terms are of the same order if $n_{1}=2s+\frac{q}{2}-p-1,$ $%
n_{2}=2s+c-p-1,$ $k=s.$

For subcritical particles the conditions for $n_{1,2}$ become

\begin{eqnarray}
n_{1} &=&2s+\frac{q}{2}-p-1, \\
n_{2} &=&2s+\frac{q}{2}-p-1+\left( s-\frac{p}{2}\right) =n_{1}+\left( s-%
\frac{p}{2}\right) .
\end{eqnarray}

As $s<\frac{p}{2},$ we see that $n_{2}<n_{1}.$ Additionally, $0<k=s<\frac{p}{%
2}.$ Thus summarizing,

\textit{for subcritical particles }$n_{1}>n_{2}$\textit{\ and }$n_{1}>\frac{q%
}{2}-p-1,$ $n_{2}>\frac{q}{2}-\frac{3p}{2}-1.$\textit{\ Additionally, }$%
0<k=s<\frac{p}{2}.$

For critical particles, 
\begin{equation}
n_{1}=\frac{q}{2}-1=n_{2}.
\end{equation}

Thus we have

\textit{for critical particles }$n_{1}=n_{2}=\frac{q}{2}-1.$\textit{\
Additionally, }$k=s=\frac{p}{2}.$

For ultracritical particles $s=\frac{p}{2},$ $c>\frac{q}{2}.$ Then, the
conditions for $n_{1,2}$ become%
\begin{eqnarray}
n_{1} &=&\frac{q}{2}-1, \\
n_{2} &=&c-1.
\end{eqnarray}

As $c>\frac{q}{2},$ we have $n_{2}>n_{1}.$ Additionally, $k=s=\frac{p}{2}.$
Thus we have

\textit{for ultracritical particles }$n_{2}>n_{1}=\frac{q}{2}-1.$\textit{\
Additionally, }$k=s=\frac{p}{2}.$

\begin{table}[tbp]
\begin{tabular}{|c|c|c|c|}
\hline
& Subcritical & Critical & Ultracritical \\ \hline
3-rd term & $n_{1}$ and $n_{2}$ are not related & $n_{1}$ and $n_{2}$ are
not related & $n_{1}$ and $n_{2}$ are not related \\ \hline
1-st and 3-rd term & $n_{1}$ and $n_{2}$ are not related & $n_{2}>n_{1}$ & $%
n_{2}>n_{1}$ \\ \hline
2-nd and 3-rd term & $n_{1}>n_{2}$ & $n_{1}>n_{2}$ & $n_{1}$ and $n_{2}$ are
not related \\ \hline
1-st, 2-nd and 3-rd term & $n_{1}>n_{2}$ & $n_{1}=n_{2}$ & $n_{2}>n_{1}$ \\ 
\hline
\end{tabular}%
\caption{ Table showing how $n_{1}$ and $n_{2}$ are related for different
types of particles and different dominant terms}
\label{tab_2}
\end{table}

\section{Analysis for different types of horizons}

Now let us investigate for what types of horizons and what types of
particles constraints arise that control the sign of $X_{s}$.

\subsection{Nonextremal horizon}

For non-extremal horizons $q=p=1.$ We will show that for all types of
particles for such horizons a finite force does not control the sign of $%
X_{s}$ $.$ As we have shown in previous Section, in all cases when a force
controls the sign of $X_{s},$ condition $k=s$ has to hold$.$ As for
subcritical particles $0<s<p/2,$ we see that $k$ has to be less than 1, what
is forbidden as we consider only integers in Taylor expansions of metric
functions. For critical and ultracritical particles situation is familiar: $%
k=s=p/2<1.$

\subsection{Extremal horizon}

For extremal horizons $q=2,$ $p\geq 2.$ In this case it is in principle
possible to have $k=s.$ Note that the 3-rd term may always be dominant if $%
k=s.$ If 1-st and 3-rd are dominant, then condition 
\begin{equation}
n_{1}=2s+\frac{q}{2}-p-1
\end{equation}

For subcritical particles this quantity is negative (because $q=2$ and $s<p/2
$) that means that the force has to diverge. But for critical and
ultracritical particles $n_{1}=0$ and force is finite.

2-nd and 3-rd terms are dominant if 
\begin{equation}
n_{2}=2s+c-p-1.
\end{equation}

For subcritical particles 
\begin{equation}
c=s+\frac{q-p}{2},\text{ \ \ }n_{2}=2s+\frac{q}{2}-p-1+\left( s-\frac{p}{2}%
\right) .
\end{equation}

We see, that as $q=2$ and $s<p/2$ for them, then $n_{2}<0.$ However, for
critical and subcritical particles $c\geq \frac{q}{2}$ and $n_{2}$ is
non-negative.

If all three terms are of the same order, all previous conditions have to
hold. To conclude, we see that in the case when a force is absent (or if it
is quite small near the horizon, so that the 3-rd term is dominant) one can
control the sign of $X_{s}$ for any type of particle. However, if force is
high enough, then it becomes possible only for critical and ultracritical
particles.

\subsection{Ultraextremal horizon}

For ultraextremal horizons ($p\geq 2$, $q\geq 3$) there is no special
restriction and all types of particles may have controllable $X_{s}.$

Thus we see that if the force is small enough (or absent) such that the
third term is dominant, then we can control the sign of $dX/dr$ for all
types of particles if the horizon is extremal or ultraextremal. If a force
becomes higher in such a way that it starts to change the dynamics of a
system (corresponding conditions for this to happen are given in the last 3
lines in Table \ref{tab_1}), we can control the sign of $dX/dr$ for
subcritical particles only if the horizon is ultraextremal, and for critical
and ultracritical only if the horizon is extremal or ultraextremal. Note
that in our previous work where we analyzed for which fine-tuned particles
and which types of horizons it is possible to have finite forces, we
obtained the same set of combinations of types of particles and types of
horizons (see Table VIII in \cite{force23}), except from the case when a
force is absent (as was mentioned after eq. (60) in \cite{force23}, such
cases were not analyzed there).

\section{Conclusions}

In this work we scrutinized self-consistent dynamics of fine-tuned particles
in the vicinity of the horizon. As it is just such types of particles is a
necessary ingredient of the BSW effect, self-consistent solutions of
equations of motion give us the condition when this effect does exist. In
doing so, we have analyzed which constraints the presence of a force imposes
on the sign of $X$ for fine-tuned particles. We have established general
relations between kinematic properties of a particle (energy, angular
momentum) and dynamic ones (forces in different directions). This has
crucial consequence for the existence (or nonexistence) of the BSW effect.
As on the horizon itself $X=0$ for fine-tuned particles, near the horizon we
examined how the sign of $dX/dr$ (or the first nonvanishing derivative on
the horizon) is related to the acceleration and properties of the metric.
Term-by-term analysis has shown that the sign of $dX/dr$ is controlled by
force only if several conditions hold: if the rate of change of the metric
coefficient $\omega $ is the same as the rate of change of $X$ near the
horizon ($k=s$) and if forces satisfy several conditions, listed in Table %
\ref{tab_1}. In addition, there is a case of freely moving particles (or
particles, for which a force near the horizon is negligibly small), for
which the relation $L_{H}\omega _{k}<0$ is to be hold$.$ In some cases,
there are no constraints on sign but there are constraints that are
necessary for $X$ to be a real quantity.

We analyzed which conditions hold for different types of particles. We have
found that for all types of particles one is led in general to control the
sign of $dX/dr$, but not for all types of horizons. For example, we have
shown that for nonextremal horizons there are no restrictions that control
the sign of $dX/dr,$ for extremal it happens only for critical and
ultracritical particles (if the force is in some sense high enough) or for
all types of particles (if the force is absent or small enough) and for
ultracritical it is irrelevant at all.

Our results show that there is a nontrivial interplay between kinematics and
dynamics in configuration when the BSW effect is possible under the action
of a force.

\end{document}